%% file: UpperLimit_eta2pp.tex
\documentclass[a4paper,11pt]{article}
\pdfoutput=0 

\usepackage{jheppub} 

\usepackage[T1]{fontenc} 

\usepackage[mathlines]{lineno}

\nopagebreak[0]

\newcommand{\mup}{\mu^+}
\newcommand{\mum}{\mu^-}
\newcommand{\pip}{\pi^+}
\newcommand{\pim}{\pi^-}
\newcommand{\pio}{\pi^0}

\title{\boldmath Upper limit on the $\eta\to\pi^+\pi^-$ branching fraction with the KLOE experiment}
\collaboration{The KLOE-2 Collaboration}

\input{author.tex}

\abstract{
Based on an integrated luminosity of 1.61 fb$^{-1}$ $e^+e^-$ collision
data collected with the KLOE detector at DA$\Phi$NE, the Frascati
$\phi$-factory, a search for  the $P$- and $CP$-violating decay 
$\eta\to\pip\pim$ has been performed. Radiative $\phi\to\eta\gamma$ 
decay is exploited to access the $\eta$ mesons. No signal is observed 
in the $\pip\pim$ invariant mass spectrum, and the upper limit on the 
branching fraction at 90\% confidence level is determined to be
${\mathcal B}(\eta\to\pip\pim)<4.9\times10^{-6}$, which is approximately 
three times smaller than the previous KLOE result. From the combination 
of these two measurements we get ${\mathcal B}(\eta\to\pip\pim) < 
4.4\times10^{-6}$ at 90\% confidence level.
}


\toccontinuoustrue

\begin{document} 
\maketitle
\flushbottom

\section{Introduction}
\label{sec:introduction}

Violation of $CP$ symmetry is a crucial ingredient in understanding 
the origin of the Baryon Asymmetry in the Universe (BAU). Although the 
Standard Model (SM) can explicitly accommodate $CP$ violation through 
a single relevant phase in the Cabibbo--Kobayashi--Maskawa (CKM) quark 
mixing matrix, this source of $CP$ violation appears largely insufficient 
to explain the observed value of the BAU.

In the SM the $P$ and $CP$ violating decay $\eta\to\pip\pim$ can proceed 
only through $CP$ violating weak interactions via a virtual $K^0_S$ meson, 
with an expected branching fraction less than $2\times10^{-27}$~\cite{PredSM}.
Introducing $CP$ violation in strong interactions through a possible 
$\theta$-term in the QCD Lagrangian~\cite{QCDtheta} would enhance this 
limit at the level of $\sim 3 \times 10^{-17}$.
Allowing additional $CP$ violation phases in the extended Higgs sector of 
the electroweak theory could generate the decay with a branching fraction 
up to $1.2\times10^{-15}$~\cite{review,Shabalin}.
By taking into account the higher-order chiral Lagrangian, the couplings
of $\eta(\eta^\prime)\pi\pi$ can be connected with the neutron electric
dipole moment (nEDM)~\cite{nEDMPred}. A recent work using the present upper 
bound on the nEDM indicates an upper limit for $\eta\to\pip\pim$ 
of $5.3\times10^{-17}$~\cite{nEDMPred2}.

A branching fraction larger than the above mentioned levels would be an 
indication of unconventional sources of $CP$ violation, which would 
possibly help solving the problem of the origin of the BAU, and making 
the search for the $\eta\to\pip\pim$ decay worth of experimental 
investigation at any accessible level at present experimental facilities.

The best upper limit to date on this branching fraction is from the KLOE 
experiment, ${\mathcal B}(\eta\to\pi^+\pi^-)<1.3\times10^{-5}$ at 90\% 
confidence level (CL)~\cite{PLB2005606-KLOEUL}, based on the analysis of 
350 pb$^{-1}$ of data collected at the $\phi$ resonance peak in years
2001 and 2002. A recent upper limit has been obtained by the LHCb 
Collaboration~\cite{PLB2017764-LHCbUL}, 
${\mathcal B}(\eta\to\pip\pim)<1.6\times10^{-5}$ at 90\% CL, searching
for the signal in $D^+\to\pi^+\eta$ and $D^+_{s}\to\pi^+\eta$ decays
produced in proton-proton collisions.
The result of a new search for the decay $\eta\to\pip\pim$ based on an 
integrated luminosity of 1.61 fb$^{-1}$ of data collected by the KLOE 
experiment in years 2004 and 2005 is reported in the following, together 
with its combination with the previous KLOE result.

\section{The KLOE detector at DA$\Phi$NE}
\label{sec:detector}

DA$\Phi$NE~\cite{DAFNE} is an $e^+e^-$ collider operated at 
center-of-mass energy of the $\phi$ meson peak, $\sim$1.020 GeV, 
with a beam energy spread of $(0.302 \pm 0.001)$ MeV. Positron and 
electron beams collide with a period of 2.7 ns at an angle of 
$\sim$25 mrad, producing $\phi$ mesons with a small transverse 
momentum $\sim13$ MeV/c. The longitudinal and horizontal width 
of the beam-beam collision region is $\Delta z\sim$12 mm and 
$\Delta x\sim$1.2 mm respectively. All these quantities
are measured run-by-run to obtain a good precision of the 
integrated luminosity~\cite{Luminosity}. The KLOE detector at 
DA$\Phi$NE is  composed of a large cylindrical drift chamber
(DC)~\cite{KLOEDC} and an electromagnetic calorimeter (EMC)~\cite{KLOEEMC}
made of lead and scintillating fibres surrounded by a superconducting coil
providing a 0.52 T axial magnetic field.
The cylindrical drift chamber, 2 m radius and 3.3 m length, is operated 
with a 90\% helium and 10\% isobutane gas mixture; its spatial resolution 
is $\sigma_{xy}\sim150~\mu$m and $\sigma_z\sim2$ mm in the transverse and
longitudinal projections, respectively. The transverse-momentum resolution
for large-angle tracks is $\sigma_{p_T}/p_T\sim0.4\%$. Vertices are
reconstructed with a spatial resolution of $\sim$ 3 mm. The calorimeter made
by lead and scintillating fibers consists of a cylindrical barrel and 
two end-caps providing a solid angle coverage of $\sim98\%$.  
The energy resolution for photons is
$\sigma_E/E=0.057/\sqrt{E(GeV)}$ and the time resolution is
$\sigma_t=54$ ps/$\sqrt{E(GeV)}\oplus100$ ps. The spatial
resolution is 1.4 cm/$\sqrt{E(GeV)}$ along the fibers and 1.3 cm in
the orthogonal direction.

The KLOE trigger system~\cite{KLOETrg} uses a two level scheme.
The first level trigger is a fast trigger with a minimal delay which
starts the acquisition of the EMC front-end-electronics. The second level
trigger is based on the energy deposits in the EMC (at least 50 MeV
in the barrel and 150 MeV in the end-caps) or on the hit multiplicity
information from the DC.
The trigger conditions are chosen to minimise the machine background,
and recognise Bhabha scattering or cosmic-ray events. Both the calorimeter
and drift chamber triggers are used for recording physical events.

The GEANFI Monte Carlo (MC)~\cite{DataHandling} simulation describes 
the geometry and material of the KLOE detector, as well as the detector 
response. Run-by-run machine background conditions are taken into account; 
the calorimeter energy deposits and drift chamber hits from beam background 
events triggered at random are overlaid onto the simulated events. 
The simulated events are processed with the same reconstruction algorithms 
as the data.
The MC production includes all the relevant $\phi$ decay channels, and
continuum processes $e^+e^-\to e^+e^-\gamma$, $\mup\mum\gamma$,
$\pip\pim\gamma$ to estimate the background contributions.
Proper scaling due to the different integrated luminosity of the samples
is taken into account when the different MC contributions are
merged together. A sample of the signal $\phi\to\eta\gamma$ with 
$\eta\to\pip\pim$ is generated to optimise the event selection criteria and
to determine the detector efficiency.

\section{Data sample and event selection}
\label{sec:eventsel}

For the selection of signal candidate events $\phi\to\eta\gamma$ 
with $\eta\to\pip\pim$, two opposite charged tracks with a vertex
near the $e^+e^-$ interaction point (IP) are required together
with an energy deposit (cluster) in the EMC compatible with the 
photon recoiling against the $\eta$ meson from the IP. 
The tracks
are reconstructed from hits in drift chamber within the polar angle
range $45^\circ<\theta<135^\circ$. The vertex is required to be within
a cylinder, 20 cm long and 8 cm of radius, centered on the IP. 
To evaluate the time of flight of particles both tracks
are required to be associated to a cluster in the EMC. 
The transverse and three-dimensional distances between the centroid 
of the associated cluster and the track extrapolation point to the 
calorimeter front surface are required to be less
than 30 cm and 100 cm, respectively (track-to-cluster association).
The cluster energy is required to be greater than 10 MeV. If there is more 
than one cluster satisfying the above requirements, the cluster with the
lowest transverse distance is assigned as the associated cluster. 
The scalar sum of the momenta of the two tracks 
must lie in the range [0.15,1.03] GeV/c.
Selected $\phi\to\rho\pi$ events are used to study
the tracking and vertex efficiencies on data and MC.
Efficiency corrections as a function of transverse and longitudinal momenta 
for charged pions reconstruction are applied to all MC samples. 
The correction for the vertex efficiency is negligible.
A background filter algorithm~\cite{DataHandling} based only on 
information from the EMC is used to reject cosmic rays and Bhabha 
scattering background events.

\begin{figure}[!htbp]
\centering
\includegraphics[width=0.45\textwidth]{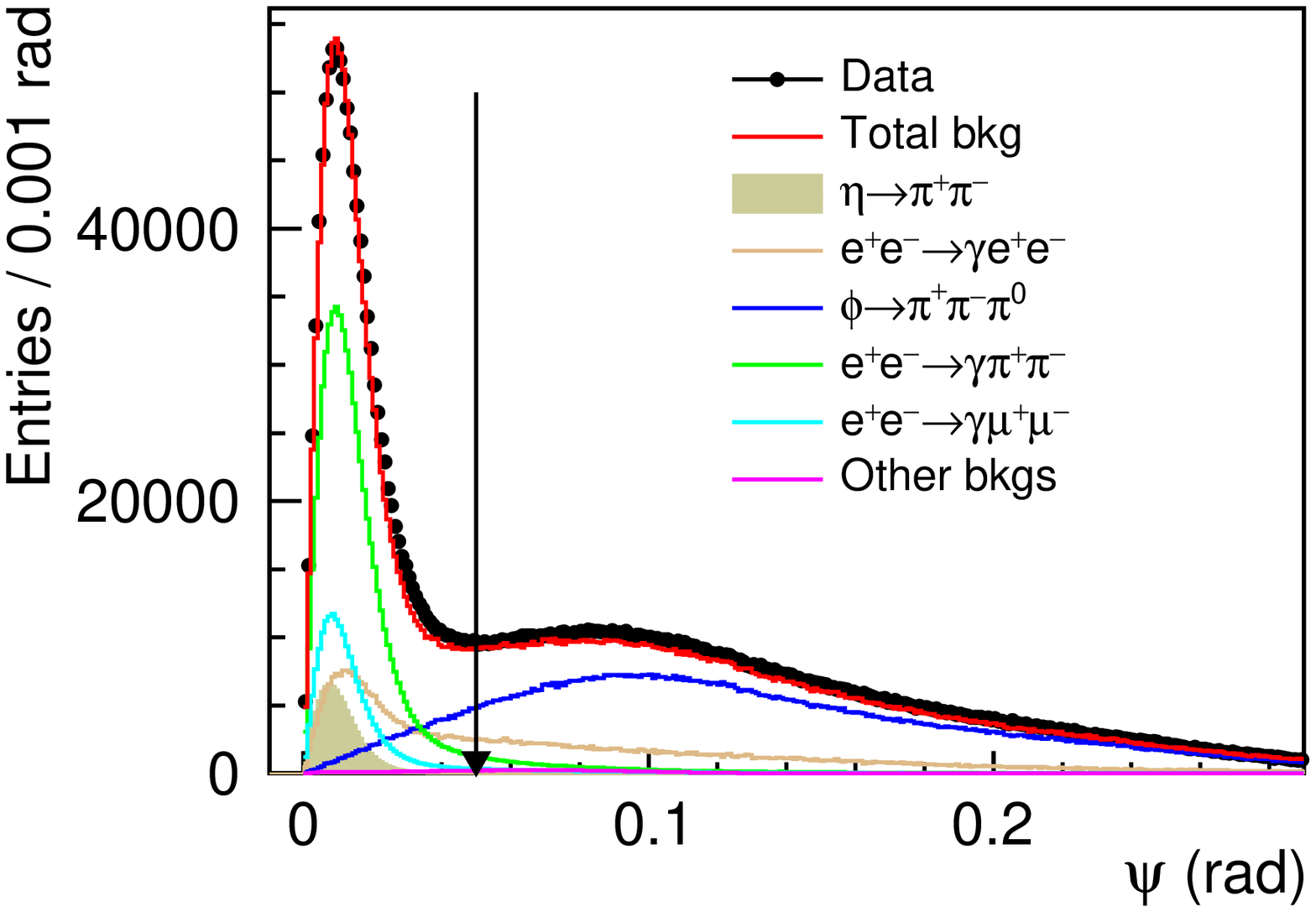}
\includegraphics[width=0.45\textwidth]{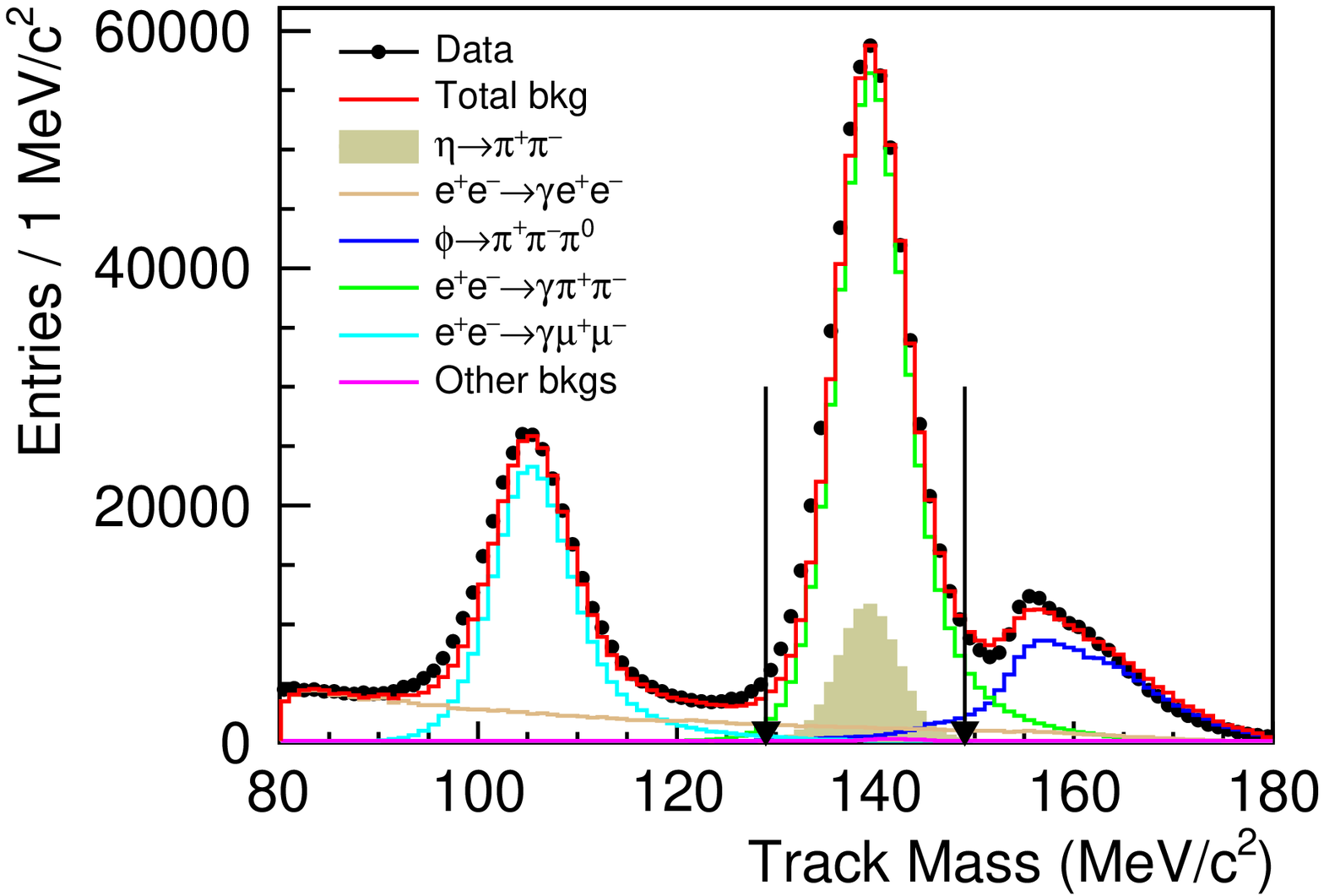}
\caption{\label{fig:omega} Left: The angle $\psi$ between the direction of 
missing momentum of $\pi^+\pi^-$ and the prompt photon. Right: Distribution 
of the mass $M_{trk}$ of the charged-particle tracks. Black dots are data; 
the red histogram is the sum of all 
background contributions evaluated from MC simulation:
$\phi\to\rho\pi$ with $\rho\to\pi\pi$ (blue histogram),
$e^+e^-\to\pip\pim\gamma$ (green histogram), 
$e^+e^-\to\mup\mum\gamma$ (cyan histogram), 
$e^+e^-\to e^+ e^-\gamma$ (yellow histogram), and the sum of other
backgrounds (violet histogram). The expected signal is shown as the
shaded histogram with the branching fraction of $\eta\to\pip\pim$ 
arbitrarily fixed to $8.8\times10^{-3}$ for visualisation purposes.
}
\end{figure}

The recoil photon candidate is selected by requiring an isolated 
energy cluster in the EMC not associated to any track.
The condition on cluster time
$|T_{cl}-R_{cl}/c|<min(5\sigma_t(E_{cl}),2ns)$ is used to 
identify a photon originating from the IP (prompt photon), where
$T_{cl}$ is the cluster time, $R_{cl}$ is the distance from the IP,
$\sigma_t$ is the energy-dependent time resolution.
To suppress background from $e^+e^-\to\pi^+\pi^-\gamma$ process, the cluster
is also required to be at large polar angle $45^\circ<\theta_\gamma<135^\circ$.
As for the two body decay $\phi\to\eta\gamma$ the recoil photon has an
energy of $363$~MeV in the $\phi$ rest frame, the selected candidate photon
is required to have an energy in the range $[250,470]$~MeV.
To match the missing energy and momentum obtained from the two tracks
with the photon kinematics, the angle $\psi$ between the direction of
the missing momentum of the two tracks and the direction of the recoil photon,
shown in the left panel of Figure~\ref{fig:omega}, is required to be less
than 0.05~rad.

\begin{figure}[!htbp]
\centering
\includegraphics[width=0.45\textwidth]{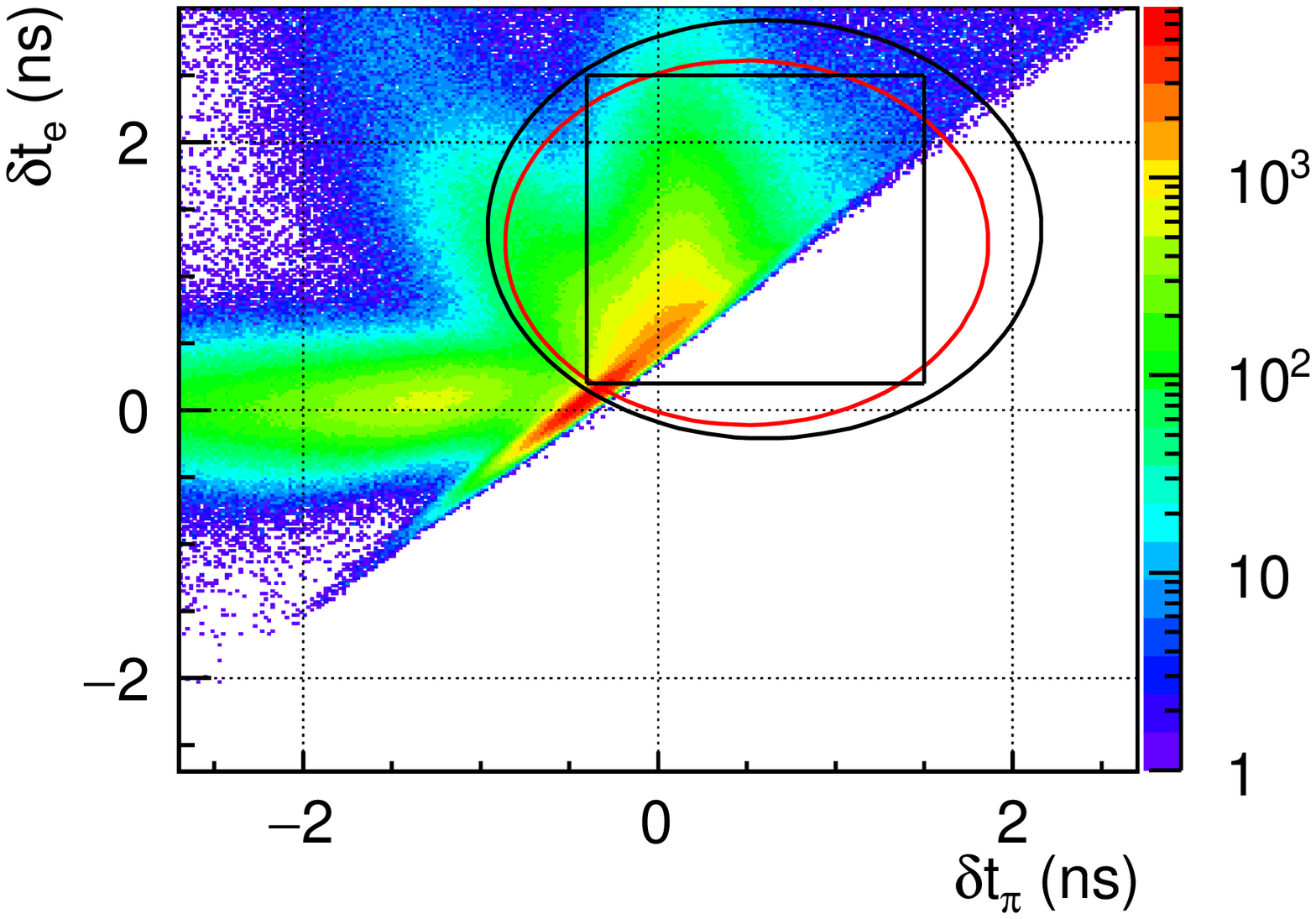}
\includegraphics[width=0.45\textwidth]{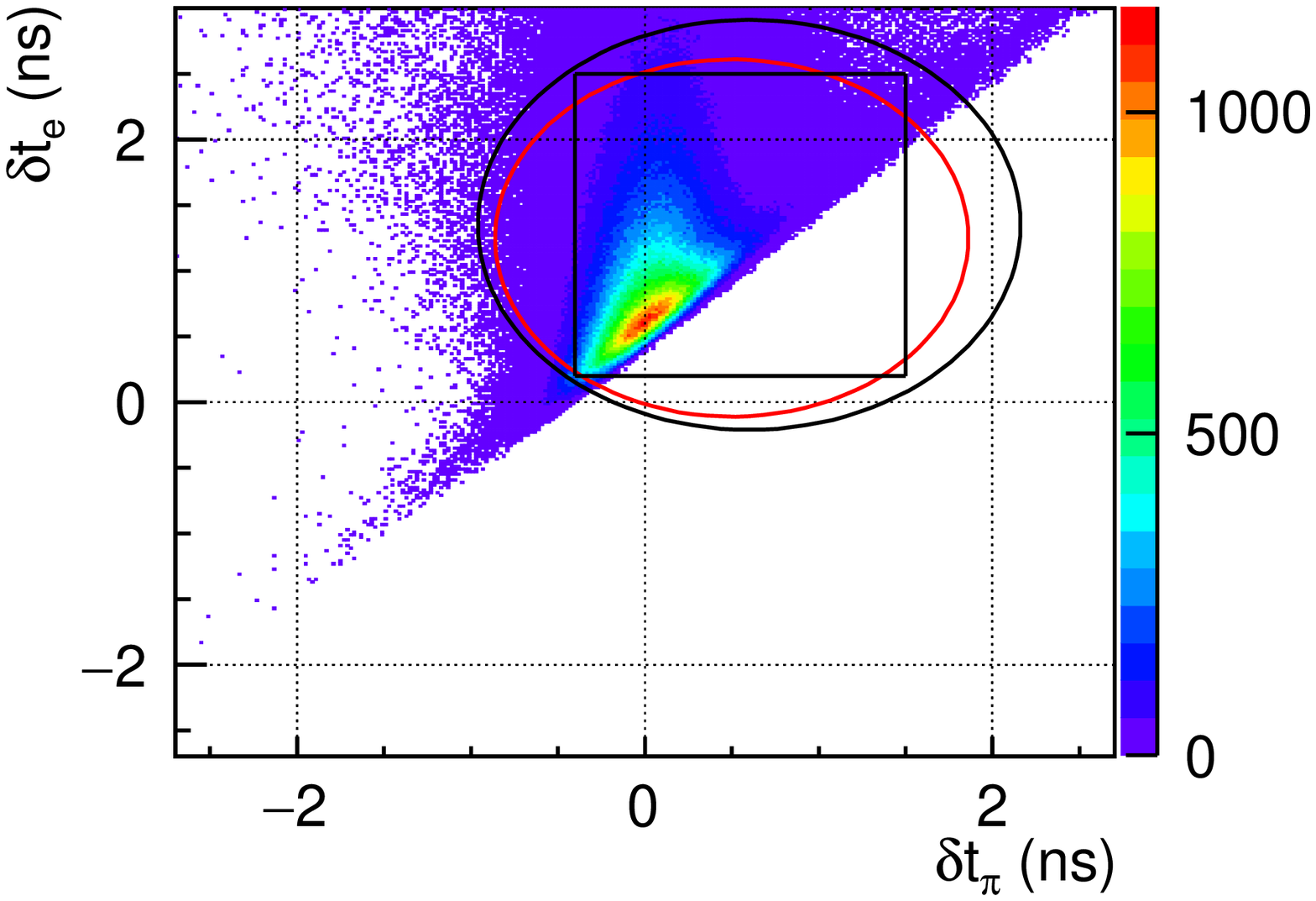}
\caption{\label{fig:tof} Scatter plot of the time difference for the pion
($\delta t_\pi$) and electron ($\delta t_e$) mass hypothesis for data (left)
and MC simulated signal (right).
Events within the rectangle are retained for further analysis,
within the elliptic shapes are for systematic uncertainty studies.
}
\end{figure}

The remaining background originates from the processes 
$e^+e^-\to e^+e^-\gamma$, $\mup\mum\gamma$, $\phi\to\rho^\pm\pi^\mp$
with $\rho^\pm\to\pi^\pm\gamma$, and $\phi\to\pip\pim\pio$ with 
an undetected photon.
To separate $\pip\pim\gamma$ and $e^+e^-\gamma$ events, particle 
identification with a time of flight technique is used. 
The difference between the time of the cluster associated to the track 
($T_{cl}$) and the time calculated from the track length $L$ and 
particle momentum $p$ under different mass hypotheses is defined as 
$\delta t_X = T_{cl}-L/(c\beta_X)$,
where $\beta_X = p/\sqrt{p^2+m_X^2}$ and $m_X$ is the pion or
electron mass, the scatter plots of $\delta t_e$ vs $\delta t_\pi$ 
for data and MC simulated signal are shown in Figure~\ref{fig:tof}.
A track with $0.2<\delta t_e<2.5$~ns and $-0.4<\delta t_\pi<1.5$~ns
is identified as a pion. 
Events with at least one pion are retained. The
cuts have been chosen to optimize the rejection of the $e^+e^-\gamma$ background,
while keeping almost unaltered efficiency on the signal.

The $\mu^+\mu^-\gamma$ and $\pip\pim\pio$ background events can be 
rejected using the mass of the charged tracks, $M_{trk}$, computed 
by assuming the $\phi$ decays to two identical mass particles and 
a photon, i.e.
\begin{equation} \label{eq:mtrk}
|\vec{p}_{\phi}-\vec{p}_{1}-\vec{p}_{2}| =
E_\phi - \sqrt{|\vec{p}_{1}|^2+M_{trk}^2} -
\sqrt{|\vec{p}_{2}|^2+M_{trk}^2},
\end{equation}
where $\vec{p}_\phi$, $\vec{p}_{1}$ and $\vec{p}_{2}$ are the three-momentum
of $\phi$ and two charged tracks, respectively; $E_\phi$ is the energy 
of the $\phi$-meson. Figure~\ref{fig:omega}-right shows the $M_{trk}$ 
distribution for data, the MC simulated signal and contributions from 
different background sources. The condition $129<M_{trk}<149$ MeV/c$^2$ 
is required to be fulfilled for candidate events to reject most of the 
backgrounds from $\mu^+\mu^-\gamma$ and $\pip\pim\pio$.

\begin{figure}[htbp]
\centering
\includegraphics[width=0.6\textwidth]{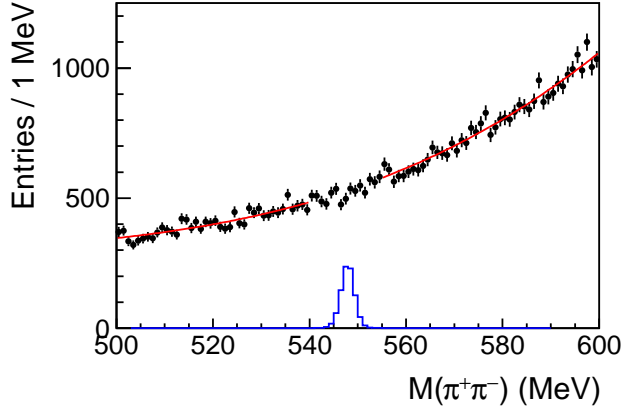}
\caption{\label{fig:Mpp_fitsb} $\pip\pim$ invariant mass distribution for 
data taken in 2004/2005. The dots with error bars are data, the red lines 
represent the fit result to $\eta$ sidebands, the blue histogram is the $\eta$ 
signal with the branching fraction arbitrarily fixed to $8.8\times10^{-5}$ for 
visualisation purposes.
}
\end{figure}

After the above selection criteria, 59,684 events remain in the 
$\eta$ mass region [500,600] MeV/c$^2$. The $\pi^+\pi^-$ invariant 
mass spectrum, $M(\pip\pim)$, is shown as the black dots in 
Figure~\ref{fig:Mpp_fitsb}, which will be used to search for the
decay $\eta\to\pip\pim$.
The survived events with a $\pip\pim\gamma$ final state are mainly from
$e^+e^-\to\pip\pim$ accompanied by initial or final state radiation,
$\phi\to f_0(980)\gamma$ with $f_0(980)\to\pip\pim$ and
$\phi\to\rho^\pm\pi^\mp$ with $\rho^\pm\to\pi^\pm\gamma$.
However, none of these backgrounds is expected to contribute as a peak
in the $\pip\pim$ invariant mass near the $\eta$ mass value.

The irreducible background in the $\eta$ signal region [540,555] MeV/c$^2$
is evaluated by performing a fit to the $\eta$ side bands, [500,540] and
[555,600] MeV/$c^2$, with a third-order polynomial function. The fit 
has $\chi^2=84.9$ with 81 degrees of freedom;
the result is illustrated by the red lines in Figure~\ref{fig:Mpp_fitsb}.
The $\eta$ signal is described by the corresponding MC simulated shape, 
shown as the blue histogram in Figure~\ref{fig:Mpp_fitsb}; to ease 
visualisation the signal branching fraction has been arbitrarily fixed 
to $8.8\times10^{-5}$.

The $M(\pip\pim)$ signal shape and resolution are validated comparing 
the $M(\pip\pim)$ distributions of data and MC for a pure sample of 
$K_S\to\pip\pim$ events.

The detection efficiency for the signal process $\phi\to\eta\gamma$
with $\eta\to\pip\pim$ is evaluated from MC to be 
$\varepsilon=(14.70\pm0.02_{stat})\%$.

\section{Upper limit on the branching fraction}
\label{sec:ulextraction}

As no evident peak is observed in the distribution of $M(\pip\pim)$
in the signal region, an upper limit on the branching
fraction of $\eta\to\pip\pim$ is extracted with the $CL_s$
technique~\cite{CLsMethod,CLsMethod2}. The $CL_s$ value is defined as
$p_{s+b}/(1-p_b)$, where $p_b$ is the $p$-value of the background only
hypothesis, $p_{s+b}$ is the $p$-value of the background plus signal
hypothesis. The procedure requires the measured $M(\pip\pim)$ spectrum, 
the shape of the estimated background, the $\eta$ signal shape, 
and the systematic uncertainties as input~\cite{CLsMethod3};
the latter will be described in Section~\ref{sec:syserr}.
It yields as output the limit on 
the branching fraction ${\mathcal B}(\eta\to\pip\pim)$, with
the number of $\eta\to\pip\pim$ signal events evaluated as follows:
\begin{equation} \label{eq:NsigCal}
 N(\eta\to\pip\pim)=N_{\eta\gamma}\cdot{\mathcal B}(\eta\to\pip\pim)\cdot
 \varepsilon.
\end{equation}
where $N_{\eta\gamma}$ is the number of $\phi\to\eta\gamma$
events with $N_{\eta\gamma} = {\mathcal L_{int}}
\cdot\sigma(e^+e^-\to\phi\to\eta\gamma)$; ${\mathcal L_{int}}$ is the
integrated luminosity, determined to be $(1.61\pm0.01)$ fb$^{-1}$ from
the very large-angle Bhabha scattering events~\cite{Luminosity}. The cross
section $\sigma(e^+e^-\to\phi\to\eta\gamma)$ has been evaluated to be
$(41.7\pm0.6)$ nb and takes into account the small variations
of $\sqrt{s}$ on a run by run basis~\cite{PhiLineShape}.

A series of statistical tests is carried out for each hypothesised
${\mathcal B}(\eta\to\pip\pim)$ to numerically evaluate the distribution 
functions for the test statistics. 
The upper limit on ${\mathcal B}(\eta\to\pip\pim)$
at 90\% CL is determined by requiring the $CL_s$ value equals
0.1 and results to be:
\begin{equation}
\mathcal{B}(\eta\to\pip\pi)<4.9\times10^{-6}.
\end{equation}
The corresponding upper limit on the number of signal events 
is $N(\eta\to\pip\pim)<48$.

\section{Systematic uncertainty}
\label{sec:syserr}

The systematic uncertainties affecting the $\phi\to\eta(\pip\pim)
\gamma$ analysis mainly originate from the non resonant backgrounds 
and the difference in variables distributions between data
and MC samples.

The estimation of non resonant backgrounds in the signal region by 
fitting the signal sidebands has a relative uncertainty of 0.5\%;
by changing the fitting function to second- or fourth-order polynomial
or varying the fitting range by $\pm$ 2 MeV/c$^2$, the result varies within
the quoted uncertainty, therefore 0.5\% is taken as
the systematic uncertainty associated with the non resonant backgrounds.

The systematic uncertainties associated to the selection criteria on
$\psi$ and $M_{trk}$ are estimated by varying the cuts within their
resolutions respectively and evaluating the corresponding effect on
$M(\pip\pim)$. The relative variation compared to the nominal
$M(\pip\pim)$ spectrum and corrected for the corresponding variation 
in the MC efficiency is taken as systematic uncertainty, which is 
determined to be 2.0\% for the $\psi$ cut, and 3.0\% for the 
$M_{trk}$ selection.

To evaluate the systematic uncertainty associated with the time of
flight selection, the lower cuts of $\delta t_e$ and $\delta t_\pi$
are varied alternatively by $\pm$ 0.1 ns. In addition, 
different selection criteria adopting elliptic cuts in the 
($\delta t_e$,~$\delta t_\pi$) plane, instead of rectangular, are checked, i.e.
$(\delta t_e-0.6)^2 + (\delta t_\pi-1.35)^2 < 1.56$ ns$^2$, or
$(\delta t_e-0.5)^2 + (\delta t_\pi-1.25)^2 < 1.36$ ns$^2$,
as shown in Figure~\ref{fig:tof}. The maximum relative
variation compared to the nominal $M(\pip\pim)$ spectrum
and corrected for the corresponding variation in the MC efficiency 
is found to be around 1.0\%, which is taken as systematic
uncertainty.

The trigger efficiency has been evaluated from the comparison of
the EMC and DC single and coincidence rates. The efficiency is
in agreement with the MC evaluation, with a negligible uncertainty.

A sample of non-filtered and non-pre-selected events, prescaled by
a factor of 20, is used to validate the efficiency of the background
filter and event pre-selection algorithms. The $e^+e^-\to\pip\pim\gamma$
process is used to estimate the differences between data and MC associated
with the two algorithms, the effect is found negligible.

All the systematic uncertainties, including the uncertainty on the
integrated luminosity~\cite{Luminosity} and the cross section
$\sigma(e^+e^-\to\phi\to\eta\gamma)$~\cite{PhiLineShape}, 
are listed in Table~\ref{tab:syserr}, where the total systematic
uncertainty is estimated as the sum in quadrature of all contributions.

\begin{table}[!htbp]
\centering
\caption{\label{tab:syserr} Summary of the systematic uncertainties.}
  \begin{tabular}{cc}\hline\hline
      Source                               & Relative uncertainty(\%)\\\hline
      Background Estimate                  &     0.5    \\
      $\psi$ cut                           &     2.0    \\
      $M_{trk}$ cut                        &     3.0           \\
      Time of flight cuts                  &     1.0           \\
      Integrated luminosity                &     0.6           \\
      $\sigma(e^+e^-\to\phi\to\eta\gamma)$ &     1.4           \\\hline
      Total                                &     4.1           \\\hline\hline
  \end{tabular}
\end{table}

\section{Combination with 2001/2002 data}
\label{sec:combination}

This section presents the procedure to combine the 2001/2002 data 
analysed in Ref.~\cite{PLB2005606-KLOEUL}
with the 2004/2005 data sample to get a combined upper limit.

In Ref.~\cite{PLB2005606-KLOEUL} 
the upper limit was determined at 90\% confidence level as:
\begin{eqnarray}
{\mathcal B}(\eta\to\pip\pim) = \frac{N(\eta\to\pip\pim)}{N_\eta \cdot \varepsilon}
<1.3\times10^{-5},
\end{eqnarray}
with $N(\eta\to\pip\pim)<33$, $\varepsilon=(16.6\pm0.2_{stat}\pm0.4_{syst})\%$, and
$N_\eta= 1.55\times10^7$ the normalisation determined from the observed
$\phi\to\eta(3\pio)\gamma$ decays with a systematic uncertainty of 2\%.

The upper limit for this totally independent data sample has been 
re-evaluated using the same procedure described in 
Section~\ref{sec:ulextraction}. The signal shape is kept the same 
as in Ref.~\cite{PLB2005606-KLOEUL}, a Gaussian function centered at 
the $\eta$ mass value $M_{\eta}=547.874$ MeV/$c^2$ measured by KLOE~\cite{KLOEEtaMass}
and a standard deviation of 1.33 MeV/$c^2$ estimated from the MC simulation. The
90\% CL upper limit on ${\mathcal B}(\eta\to\pip\pim)$ is determined
to be $1.36\times10^{-5}$, which is consistent with the published
result. 

The procedure described in Section~\ref{sec:ulextraction} is then 
used to evaluate the upper limit combining the two data samples
taking into account their differences in the $\eta$ signal shape,
the observed $M(\pip\pim)$ spectra and the shape of the estimated 
backgrounds, similarly to the procedure used in Ref.~\cite{UbosonComb}. 
The systematic uncertainties estimated for both 2001/2002 and 2004/2005 
samples are given as input to the procedure. The resulting upper limit 
at 90\% CL is:
\begin{equation}
\mathcal{B}(\eta\to\pip\pim)<4.4\times10^{-6}
\end{equation}
which is almost a factor of three smaller than the previous limit.

\acknowledgments

We warmly thank our former KLOE colleagues for the access to the data
collected during the KLOE data taking campaign. We thank the DA$\Phi$NE
team for their efforts in maintaining low background running conditions
and their collaboration during all data taking. We want to thank our
technical staff: G.~F.~Fortugno and F.~Sborzacchi for their dedication
in ensuring efficient operation of the KLOE computing facilities;
M.~Anelli for his continuous attention to the gas system and detector safety;
A.~Balla, M.~Gatta, G.~Corradi and G.~Papalino for electronics maintenance;
C.~Piscitelli for his help during major maintenance periods. This work was
supported in part by the Polish National Science Centre through the Grants
No. 2013/11/B/ST2/04245, 2014/14/E/ST2/00262, 2014/12/S/ST2/00459,
2016/21/N/ST2/01727, 2016/23/N/ST2/01293, 2017/26/M/ST2/00697.

\end{document}

%% file: author.tex
\newcommand{\affuni}[2]{Dipartimento di Fisica dell'Universit\`a #1, #2, Italy.}
\newcommand{\affinfn}[2]{INFN Sezione di #1, #2, Italy.}

\author[d]{D.~Babusci}
\author[v]{M.~Berlowski}
\author[d]{C.~Bloise}
\author[d]{F.~Bossi}
\author[s]{P.~Branchini}
\author[r,s]{A.~Budano}
\author[u]{B.~Cao}
\author[r,s]{F.~Ceradini}
\author[d]{P.~Ciambrone}
\author[a,d]{F.~Curciarello}
\author[c]{E.~Czerwi\'nski}
\author[n,o]{G.~D'Agostini}
\author[d]{E.~Dan\`e}
\author[n,o]{V.~De~Leo}
\author[d]{E.~De~Lucia}
\author[d]{A.~De~Santis}
\author[d]{P.~De~Simone}
\author[r,s]{A.~Di~Cicco}
\author[n,o]{A.~Di~Domenico}
\author[d]{D.~Domenici}
\author[d]{A.~D'Uffizi}
\author[p,q]{A.~Fantini}
\author[d]{P.~Fermani}
\author[t,o]{S.~Fiore}
\author[c]{A.~Gajos}
\author[n,o]{P.~Gauzzi}
\author[d]{S.~Giovannella}
\author[s]{E.~Graziani}
\author[g,h]{V.~L.~Ivanov}
\author[u]{T.~Johansson}
\author[d, w]{X.~Kang}
\emailAdd{xiaolin.kang@lnf.infn.it}
\author[c]{D.~Kisielewska-Kami\'nska}
\author[g,h]{E.~A.~Kozyrev}
\author[v]{W.~Krzemien}
\author[u]{A.~Kupsc}
\author[g,h]{P.~A.~Lukin}
\author[f,b]{G.~Mandaglio}
\author[d,m]{M.~Martini}
\author[p,q]{R.~Messi}
\author[d]{S.~Miscetti}
\author[d]{D.~Moricciani}
\author[c]{P.~Moskal}
\author[c]{S.~Parzych}
\author[s]{A.~Passeri}
\author[l,o]{V.~Patera}
\author[n,o]{E.~Perez~del~Rio}
\author[d]{P.~Santangelo}
\author[j,k]{M.~Schioppa}
\author[r,s]{A.~Selce}
\author[c]{M.~Silarski}
\author[d,e]{F.~Sirghi}
\author[g,h]{E.~P.~Solodov}
\author[s]{L.~Tortora}
\author[i]{G.~Venanzoni}
\author[v]{W.~Wi\'slicki}
\author[u]{M.~Wolke}

\affiliation[a]{Dipartimento di Fisica e Astronomia "Ettore Majorana",
Universit\`a di Catania, Italy}
\affiliation[b]{\affinfn{Catania}{Catania}}
\affiliation[c]{Institute of Physics, Jagiellonian University, Cracow, Poland.}
\affiliation[d]{Laboratori Nazionali di Frascati dell'INFN, Frascati, Italy.}
\affiliation[e]{Horia Hulubei National Institute of Physics and Nuclear Engineering, M\v{a}gurele, Romania}
\affiliation[f]{Dipartimento di Scienze Matematiche e Informatiche, Scienze Fisiche e Scienze della Terra dell'Universit\`a di Messina, Messina, Italy.}
\affiliation[g]{Budker Institute of Nuclear Physics, Novosibirsk, Russia.}
\affiliation[h]{Novosibirsk State University, Novosibirsk, Russia.}
\affiliation[i]{\affinfn{Pisa}{Pisa}}
\affiliation[j]{\affuni{della Calabria}{Rende}}
\affiliation[k]{INFN Gruppo collegato di Cosenza, Rende, Italy.}
\affiliation[l]{Dipartimento di Scienze di Base ed Applicate per l'Ingegneria dell'Universit\`a ``Sapienza'', Roma, Italy.}
\affiliation[m]{Dipartimento di Scienze e Tecnologie applicate, Universit\`a ``Guglielmo Marconi", Roma, Italy.}
\affiliation[n]{\affuni{``Sapienza''}{Roma}}
\affiliation[o]{\affinfn{Roma}{Roma}}
\affiliation[p]{\affuni{``Tor Vergata''}{Roma}}
\affiliation[q]{\affinfn{Roma Tor Vergata}{Roma}}
\affiliation[r]{Dipartimento di Matematica e Fisica dell'Universit\`a 
``Roma Tre'', Roma, Italy.}
\affiliation[s]{\affinfn{Roma Tre}{Roma}}
\affiliation[t]{ENEA, Department of Fusion and Technology for Nuclear Safety and Security, Frascati (RM), Italy}
\affiliation[u]{Department of Physics and Astronomy, Uppsala University, Uppsala, Sweden.}
\affiliation[v]{National Centre for Nuclear Research, Warsaw, Poland.}
\affiliation[w]{School of Mathematics and Physics, China University of Geosciences (Wuhan), Wuhan, China.}